\documentclass{elsart}
\usepackage{graphicx}
\usepackage{amssymb}

\begin{document}
\begin{frontmatter}

\title{Rotation-driven prolate-to-oblate shape phase transition
in $^{190}$W: A projected shell model
study\protect\footnote{Dedicated to the memory of Professor Dr.
Hans-J\"org Mang.}}

\author[1,2,3]{Yang Sun},
\author[4]{Philip M. Walker},
\author[5,7]{Fu-Rong Xu},
\author[6,7]{Yu-Xin Liu}

\address[1]{Institute of Modern Physics, Chinese Academy of
 Sciences, Lanzhou 730000, P. R. China}
\address[2]{Department of Physics, Shanghai Jiao Tong University,
 Shanghai 200240, P. R. China}
\address[3]{Joint Institute for Nuclear Astrophysics,
 University of Notre Dame, Notre Dame, IN 46556, USA}
\address[4]{Department of Physics, University of Surrey,
Guildford GU2 7XH, United Kingdom}
\address[5]{Department of Technical Physics, Peking
University, Beijing 100871, P. R. China}
\address[6]{Department of Physics, Peking University, Beijing 100871,
P. R. China}
\address[7]{Center of Theoretical Nuclear Physics, National Laboratory
of Heavy Ion Accelerator, Lanzhou 730000, P. R. China}

\date{\today}

\begin{abstract}
A shape phase transition is demonstrated to occur in $^{190}$W by
applying the Projected Shell Model, which goes beyond the usual
mean-field approximation. Rotation alignment of neutrons in the
high-$j$, $i_{13/2}$ orbital drives the yrast sequence of the
system, changing suddenly from prolate to oblate shape at angular
momentum 10$\hbar$. We propose observables to test the picture.
\end{abstract}

\begin{keyword}
Prolate and oblate shapes \sep Shape phase transition \sep Rotation
alignment \sep Projected Shell Model

\PACS 21.10.Re, 21.60.Cs, 27.80.+w
\end{keyword}
\end{frontmatter}

\newpage

Understanding the fundamental excitations of many-fermion systems
and transitions between different excitation modes is an important
issue in mesoscale physics. In nuclei where the ground state is
typically a superconducting state with paired nucleons moving in
orbits, the low-lying excitation spectrum is generally formed by
collective motion (for example, rotation and vibration with
different nuclear shapes, or deformation) and nucleon pair
breaking. Being a finite-size quantum system, for certain numbers
of protons and neutrons, a subtle rearrangement of only a few
nucleons among the orbitals near the Fermi surface can result in
completely different collective modes, constituting a suitable
situation for studying the shape phase transition
\cite{Casten06,Iachello98,Iachello04}.

There is evidence for several different types of shape phase
transition in nuclei. Casten {\it et al.} \cite{Casten99}
systematically studied spherical-to-deformed ground-state phase
transitions as a function of neutron number and proton number; and
Regan {\it et al.} \cite{Regan03} observed the evolution from
collective vibrational to rotational structure as a function of
angular momentum (see also theoretical discussions
\cite{Jolie04,Liu06}). What we discuss in this Letter is a robust
example of a shape phase transition that occurs along an yrast
cascade (the locus of lowest energy states of each spin) between
states of prolate and oblate shape in an isolated nucleus.

The neutron-rich Hf and W nuclei with neutron number $N\approx
116$ are expected to exhibit interesting critical-point phenomena
due to competing collectivity of prolate and oblate shapes. Since
experimental evidence is sparse, we refer mainly to theoretical
predictions (e.g. Refs. \cite{Xu00,Stevenson05}). In these nuclei,
the ground state and low-spin states are of a prolate shape.
However, the occupation of low-$K$, $i_{13/2}$ neutron intruder
orbits when the shape is oblate favors rotation alignment at
moderate spin values. ($K$ is the symmetry-axis projection of the
angular-momentum vector.) This was suggested by energy-surface
calculations for $^{190}$W based on a mean-field method
\cite{WX06a}. Ref. \cite{WX06a} also suggested that the $^{190}$W
structure is a candidate for a classic case of ``giant
backbending", of the type originally described by Hilton and Mang
\cite{Hilton79} in the 1970's. Confirmation of these suggestions
requires additional theoretical investigations as well as
experimental data. Here we show, with aid of detailed
shell-model-type calculations, that $^{190}$W can exhibit an
excellent example of a shape phase transition that differs from
the previously known types of phase transition reported in the
literature \cite{Casten99,Regan03}. The combination of electrical
quadrupole moments or B(E2) values, and gyromagnetic ratios (g
factors), should be able to give firm confirmation. The isomerism
\cite{WD99,w190} is a key feature in giving access to these
observables.

Phase transitions in nuclei have been theoretically described
mainly by means of algebraic models \cite{Jolie04,Liu06,Frank06},
in which a transition-driving control parameter appears explicitly
in the Hamiltonian. There have been microscopic studies based on
mean-field theories (for example, Ref. \cite{Lala06}). Using the
Landau theory, Alhassid {\it et al.} \cite{ALZ86} found a
universal behavior of rapid changes of the equilibrium shape from
the prolate ground state to oblate excited states in hot rotating
nuclei. However, the nuclear shell model can provide a more
fundamental basis to study phase transitions. The advantage of a
shell-model study is that one may see the microscopic origin of a
phase transition by analyzing wave functions. The early work of
Federman {\it et al.} \cite{Pittel79} showed perhaps the first
example of such a kind. The more recent phase transition study
with the Monte Carlo Shell Model \cite{Otsuka01} and large-scale
shell-model calculations \cite{Hase07} are other examples.
However, these shell models are applicable only to small nuclear
systems, and $^{190}$W is far too large for meaningful
calculations of this type.

\begin{figure*}
\includegraphics[width=13cm]{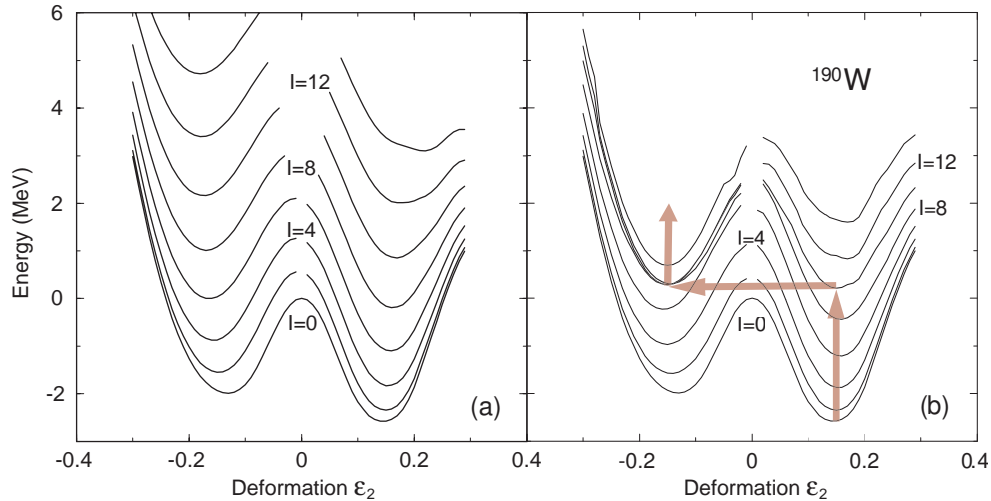}
\caption{Energy surface calculation for $^{190}$W. (a) Energies
obtained with projection on qp vacuum states only. (b) Energies
from calculation with basis that includes 2-qp states.}
\label{fig1}
\end{figure*}

The Projected Shell Model (PSM) \cite{PSM} is a shell model that
uses deformed bases and the projection technique. It is applicable
to any large size of deformed systems (with extreme examples from
superdeformed \cite{Sun97} to superheavy nuclei \cite{Rodi06}). The
PSM's two-body residual interactions are of the quadrupole plus
pairing type, with the quadrupole-pairing term included \cite{PSM}.
Wave functions of the PSM are written in terms of
angular-momentum-projected multi-quasiparticle (qp) states
\begin{equation}
|\psi^{I}_M\rangle = \sum_{\kappa} f^{I}_{\kappa} \hat
P^{\,I}_{MK_\kappa}|\phi_\kappa\rangle , \label{wf}
\end{equation}
where the index $\kappa$ labels basis states.  $\hat P^{\,I}_{MK}$
is the angular-momentum-projection operator \cite{PSM} and the
coefficients $f^{I}_{\kappa}$ are weights of the basis states. For
even-even nuclei, $|\phi_\kappa\rangle$ in Eq. (\ref{wf}) is
\begin{equation}
\{ |0\rangle, a^\dagger_\nu a^\dagger_\nu |0\rangle, a^\dagger_\pi
a^\dagger_\pi |0\rangle, \cdots \} \label{space}
\end{equation}
where $a^\dagger_\nu$ and $a^\dagger_\pi$ are the creation operator
for neutrons and protons, respectively. The qp states are obtained
from a deformed Nilsson calculation (with the Nilsson parameters
taken from Ref. \cite{BR85}) followed by a BCS calculation, in a
model space with three major shells for each kind of nucleon ($N$ =
4, 5, 6 for neutrons and $N$ = 3, 4, 5 for protons). The
corresponding qp vacuum is $|0\rangle \equiv |\varepsilon\rangle$ at
deformation $\varepsilon$. We assume axial symmetry in the basis
(consistent with Ref.~\cite{WX06a}) and, therefore, each basis state
in (\ref{space}) can be labeled by the $K$ quantum number. While the
basis states have good $K$, the states of Eq. (\ref{wf}) generally
do not, as they are linear combinations of various $K$-states.

Fig. 1 shows the projected energy surface calculation for
$^{190}$W. The quantities plotted in Fig. 1 are
\begin{equation}
E^I(\varepsilon) = \langle \psi^{I}(\varepsilon)| \hat H
|\psi^{I}(\varepsilon)\rangle .
\end{equation}
These are energies with different angular momenta $I=0,2,\cdots$
calculated as a function of basis deformation $\varepsilon$,
varying from negative values (corresponding to oblate shapes) to
positive values (corresponding to prolate shapes). The energy
curves show that there are two pronounced minima, sitting
respectively at the prolate and oblate side, with
$|\varepsilon|\approx 0.15$. The minimum at $\varepsilon\approx
+0.15$ is lower, suggesting that the ground state of $^{190}$W is
of prolate shape, which is consistent with the ground state
deformation obtained in Refs. \cite{Stevenson05,WX06a}. The
excited $I=0$ state, i.e. the minimum at the oblate side with
$\varepsilon\approx -0.15$, is about 0.6 MeV higher than the
ground state. Thus the result predicts a competing picture of
prolate and oblate shapes with two sets of corresponding
collective states. Dynamic perturbations may change the balance in
the competition as the order parameter varies, causing a
prolate-to-oblate phase transition.

Perturbations in quantum systems can arise from different sources.
The dynamic process responsible for the phase transition in this
case is the nuclear rotation and the order parameter is the total
spin $I$. As the nucleus rotates, individual nucleon pairs tend to
align their spins with the collective rotation axis through the
Coriolis force. The alignment lowers the system's energy; the
amount of energy gain depends primarily on the microscopic
structure of the aligned particles. To see this clearly, we
compare two calculations in Fig. 1. The left figure shows the
calculation in which $|\psi^{I}\rangle$ contains the projected qp
vacuum state only and all 2-qp states in basis (\ref{space}) are
switched off. The underlying physics is that there is no rotation
alignment allowed in the calculation. In such a case, minima on
the prolate side remain lower for every state. This implies that
the prolate shape always wins in the competition with the oblate
shape if no rotation alignment can happen.

Calculations that include 2-qp states in the model space are shown
in the right figure in Fig. 1. With inclusion of 2-qp states, energy
levels of higher spins are compressed considerably. As the arrows
indicate, along the yrast cascade, the path for the lowest state at
each $I$ starts from the prolate ground state $I=0$, goes up till
$I=10$, and then jumps to the oblate side. The wave functions with
prolate and oblate shapes are expected to be very different, with a
very small overlap between them. Therefore, the above result implies
that the yrast cascade of $\gamma$-rays is broken and the $\gamma$
transition between the oblate  and prolate states may be strongly
hindered. This is a favorable situation for the oblate state at the
phase transition to be isomeric. We emphasize that this type of
calculation contains beyond-mean-field correlations in the sense
that each point in the curves in Fig. 1 is obtained by projecting
the intrinsic configurations (deformed mean-field results) onto
states with good angular momentum and mixing the projected states
through residual interactions.

\begin{figure*}
\includegraphics[width=12cm]{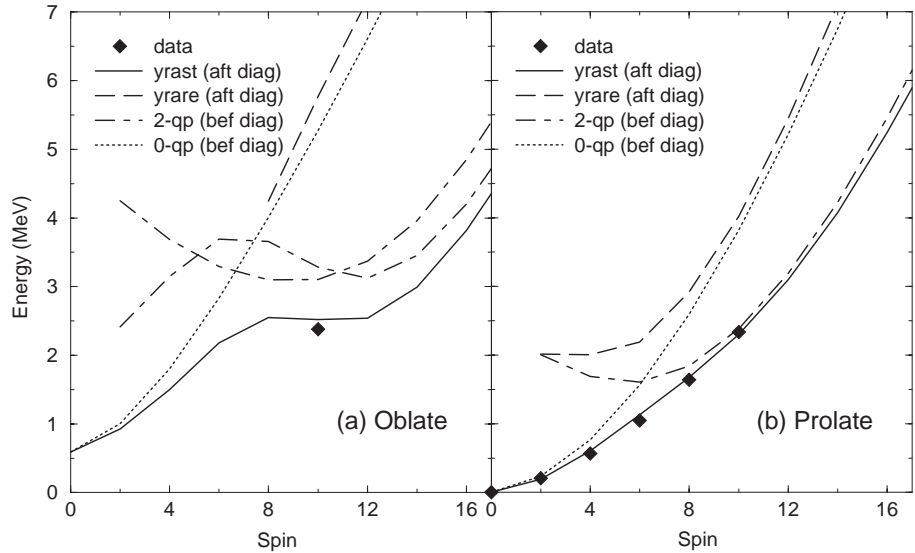}
\caption{Selected bands in the $^{190}$W calculation. (a) Bands
with oblate shape (calculated at $\varepsilon=-0.15$). (b) Bands
with prolate shape (calculated at $\varepsilon=+0.15$).}
\label{fig2}
\end{figure*}

A phase transition in a quantum many-body system generally refers
to an abrupt, qualitative change in the wave function. This
subject is of significant interest for many subfields, and is an
attractive current topic in nuclear physics
\cite{Casten06,Iachello98,Iachello04}. Our results in Fig. 1 show
a clear example of a shape phase transition occurring in an
isolated nucleus between two deformed (prolate and oblate) shapes
driven by the $i_{13/2}$ rotation alignment. Coexistence of
near-spherical and deformed shapes is a known effect for nuclei
near the proton shell closure (see for example $^{186}$Hg
\cite{Hg186}). However, the present case with $N = 116$, which
goes further to the neutron-rich region and has additional valence
protons, is more robust. In $^{186}$Hg the ground band is close to
spherical and not well developed, but there is also a spin-16 band
crossing. This latter feature has parallels with the behavior of
$^{190}$W, where the prolate (oblate) wave function contains the
highest (lowest) $K$-components of the neutron $i_{13/2}$ orbit
(see later discussions), and the wave functions before and after
the transition are therefore sharply different.

As long as a nucleus exhibits well-defined deformation minima, such
as those shown in Fig. 1, one can apply the PSM and perform detailed
shell-model analysis at the minima. To further understand the
results in Fig. 1, we show several representative bands in Fig. 2.
Angular-momentum-projection on a multi-qp state
$|\phi_\kappa\rangle$ with a sequence of $I$ generates a band. One
may define the rotational energy of a band (band energy) using the
expectation values of the Hamiltonian with respect to the projected
$|\phi_\kappa\rangle$ \cite{PSM}
\begin{equation}
E^I_\kappa = {{\langle \phi_\kappa |\hat H \hat P^I_{K_\kappa
K_\kappa}|\phi_\kappa\rangle}\over{\langle \phi_\kappa | \hat
P^I_{K_\kappa K_\kappa}|\phi_\kappa\rangle}}.
\end{equation}
These are bands before configuration mixing (diagonal elements in
the Hamiltonian matrix) and are denoted as ``bef diag" in Fig. 2.
Those after configuration mixing, i.e. the superposition of all
basis states $\kappa$ in (\ref{wf}), are denoted as ``aft diag".
Bands after mixing are solutions of the eigenvalue problem and
dynamics is thus taken into account.

In the calculation for Fig. 2, shell model diagonalization is
carried out at two deformations, $\varepsilon=\pm 0.15$ (see Fig.
1). Note that in the calculation, about 100 2-qp states are
included in each diagonalization. We present only a few of them to
illustrate the physics. We recall that in $^{190}$W, the neutron
occupation is very different at the two energy minima. On the
prolate side, one can find high-$K$ orbits such as ${9\over
2}[624]$ and ${{11}\over 2}[615]$ close to the neutron Fermi
surface, while on the oblate side, only the lowest $K$-states of
the $i_{13/2}$ intruder shell appear around the Fermi surface. In
the right part of Fig. 2, the ground band (dotted curve) starts
from the origin and goes up monotonically. At $I=6$, the $K=1$,
2-qp band (dotted-dashed) built by ${9\over 2}[624]\oplus
{{11}\over 2}[615]$ quasi-neutrons crosses the ground band.
However, interactions at the crossing region are so strong that
they repel the crossing bands from each other, forming two smooth
sequences. In fact, the resulting bands (solid and dashed curves)
are smooth bands as if no band crossing had occurred. The lowest
states (solid curve) obtained after configuration mixing reproduce
very well the known data (filled diamonds) \cite{w190}. On the
other hand, in the left part of Fig. 2, the behavior of the
low-$K$ 2-qp states on the oblate side, ${1\over 2}[660]\oplus
{{3}\over 2}[651]$ and ${1\over 2}[660]\oplus {{5}\over 2}[642]$
(both dotted-dashed), is quite different from the 2-qp state on
the prolate side. After $I=6$, the combined behavior of these two
bands gives rise to nearly degenerate states at $I=8, 10$, and 12
lying lowest in energy. These configurations are predicted to be
dominant in the lowest oblate states (solid curve), which compares
well with the tentative data point from Ref. \cite{w190}.

One can thus understand that the compression of the states with
$I=8, 10$, and 12 of the oblate shape is a consequence of the
particular rotation-alignment effect of the low-$K$, $i_{13/2}$
neutrons. The effect considerably lowers the total energy, which
helps the oblate states to eventually win the prolate-oblate
competition. As a result, a prolate-to-oblate shape phase
transition occurs along the yrast cascade, and the oblate state
with $I=10$ becomes isomeric.

Though not shown in Fig. 2, we have found several 2-qp high-$K$
bands with prolate shape in the calculation. The bandhead energies
are ranging down to 1.2 MeV of excitation. As some of them lie low
in energy and have $K$-quantum numbers much differing from the
$K=0$ ground band, these high-$K$ bands are also predicted to be
$K$-isomers. The dominant structures of the lowest such bands are:
$K^\pi=6^+, {3\over 2}[512]\oplus {{9}\over 2}[505]$; $K^\pi=10^+,
{9\over 2}[624]\oplus {{11}\over 2}[615]$; $K^\pi=7^-, {3\over
2}[512]\oplus {{11}\over 2}[615]$; and $K^\pi=10^-, {9\over
2}[505]\oplus {{11}\over 2}[615]$.

The energy minimum with oblate shape (see Fig. 1), which lies
about 0.6 MeV above the ground state, is a $0^+$ shape isomer
predicted in our calculation. A $0^+$ shape isomer has been
observed in $^{72}$Kr \cite{Kr72}. A recent experiment at GSI may
have found evidence \cite{Regan07} for $\beta$-decay to this shape
isomer.

\begin{figure}
\includegraphics[width=6.5cm]{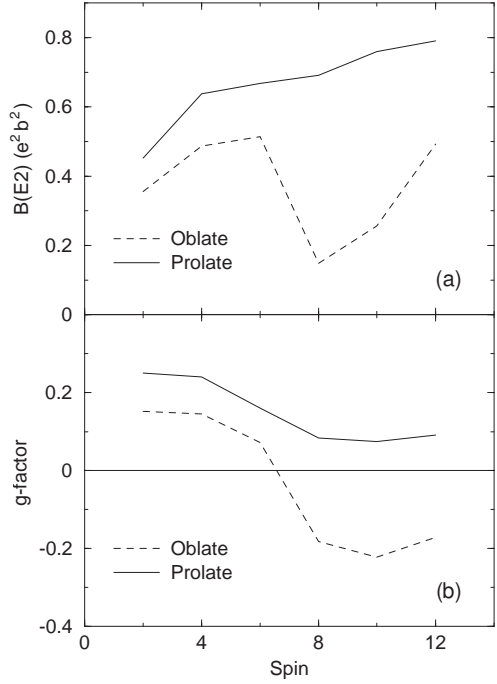}
\caption{Predicted electromagnetic properties for $^{190}$W. (a)
B(E2, $I\rightarrow I-2$) values. (b) $g$ factors.} \label{fig3}
\end{figure}

A crucial test for the above predictions is to measure additional
observables for the states before and after the phase transition.
As the states belong to very different structures, one may find
clear indications when studying their electromagnetic properties,
e.g. B(E2) values and g-factors. B(E2) values that are related to
the electric quadrupole transition probability from an initial
state $I$ to a final state $I-2$ are given by
\begin{equation}
B(E2, I\rightarrow I-2) = \frac {e^2}{2I + 1} \left| \right<
\psi^{I-2} || \hat {Q}_2 ||\psi^I \left> \right|^2. \label{B2E2}
\end{equation}
The effective charges used in the calculation are the standard
ones: $e_\pi=1.5e$ and $e_\nu=0.5e$. Magnetic properties are
described through $g$ factors, which are defined as
\begin{equation}
g(I) = \frac {\langle \psi^I || \hat \mu ||\psi^I \rangle}{\mu_N
\sqrt{I(I+1)}}. \label{g}
\end{equation}
In the $g$-factor calculation, we use the standard values for
$g_l$ and $g_s$, namely, the free values for $g_l$ and the free
values damped by a 0.75 factor for $g_s$. In Eqs. (\ref{B2E2}) and
({\ref{g}), wave functions $\left|\psi^I\right>$ are those of Eq.
(\ref{wf}).

Presented in the upper part of Fig. 3 are two sets of B(E2)
values, calculated at the prolate and oblate minima. It can be
seen that the values at the oblate minimum are clearly predicted
to be smaller than those at the prolate minimum for the spin
states $I\le 6$. A drastic drop of the oblate B(E2) occurs at
$I=8$, due to the sharp band crossing of the low-$K$ $i_{13/2}$
neutrons. Thus we predict very different B(E2) values (and hence
also quadrupole moments) for the prolate and oblate structures.
Similar conclusions can be drawn for $g$ factors. As the lower
part of Fig. 3 shows, there are significant differences between
the prolate and oblate $g$ factors. They both show a drop at $I=6$
and 8, due to the rotation alignment of neutrons; however a much
greater drop occurs for the oblate $g$ factors, which even causes
the sign to change. Experimental observation of the differences in
electromagnetic properties will definitely determine the structure
around the crossing point, and thus provide a crucial test for our
prediction. We note that new experimental information may soon be
available \cite{Regan07}.

In conclusion, by using the Projected Shell Model, which conserves
total angular momentum and includes configuration mixing beyond
the usual mean-field, we have demonstrated a new type of shape
phase transition that occurs along the yrast cascade in $^{190}$W.
The transition is driven by the dynamic process of rotation
alignment of the high-$j$ neutrons, which leads to the preferred
state suddenly changing shape from prolate to oblate at $I=10$.
The transition results in oblate shape isomers, extending the
predictions of Ref. \cite{WX06a}. We have proposed testable
quantities for future experiments to measure. Finally, we note
that our searching for energy minima has been restricted to axial
symmetry. A more general case will be minimization in the
$\varepsilon$-$\gamma$ plane with three-dimensional angular
momentum projection. This is the subject of future work.

This work was supported in part by the U.S. NSF through grant
PHY-0216783, the UK EPSRC and AWE plc., the NNSF of China under
contract Nos. 10425521, 10475002, 10525520, 10575004, 10675007, the
Key Grant Project of Chinese Ministry of Education (CMOE) under
contact No. 305001, the Research Fund for the Doctoral Program of
Higher Education of China through grant 20040001010, and by the
Chinese Major State Basic Research Development Program through grant
2007CB815005.



\end{document}